\documentclass{article}
\usepackage{geometry}                
\usepackage{graphicx}
\usepackage{epstopdf}
\usepackage{amsmath}
\usepackage{amsfonts}
\usepackage{amssymb}
\usepackage{amsthm}

\usepackage{url}

\newtheorem{proposition}{Proposition}

\newtheorem{theorem}[proposition]{Theorem}

\begin{document}

\title{Using discrete Darboux polynomials to detect and determine preserved measures and integrals of rational maps}

\author{E Celledoni$^1$, C Evripidou$^2$, D I McLaren$^2$, \\
B Owren$^1$, G R W Quispel$^2$, B K Tapley$^1$ and P H van der Kamp$^2$\\[5mm]
$^1$ 	Department of Mathematical Sciences,
	NTNU,
	7491 Trondheim,
	Norway\\
$^2$ 	Department of Mathematics,
	La Trobe University,
	Bundoora, VIC 3083, Australia   
}

\maketitle

\begin{abstract}
\noindent
In this Letter we propose a systematic approach for detecting and calculating preserved measures and integrals of a rational map. The approach is based on the use of cofactors and Discrete Darboux Polynomials and relies on the use of symbolic algebra tools. Given sufficient computing power, all rational preserved integrals can be found.
 We show, in two examples, how to use this method to detect and determine preserved measures and integrals of the considered rational maps.
\end{abstract}

\section{Introduction}
The search for preserved measures and integrals of ordinary differential equations (ODEs) has been at the forefront of mathematical physics since the time of Galileo and Newton.

\medskip \noindent In this Letter our aim will be to develop an analogous theory for the (arguably more general) discrete-time case. This will lead to essentially linear algorithms for detecting and determining preserved measures and first and second integrals of (discrete) rational maps (both integrable and non-integrable).

\medskip \noindent But before we consider the discrete case, let us look at the continuous case, i.e. ODEs.

\medskip \noindent  Consider two polynomials $P_1$ and $P_2$:
\begin{eqnarray*}
P_1(\mathbf{x}) &=& \sum a_{i_1,\dots,i_n} x_1^{i_1} \dots x_n^{i_n}  \\
P_2(\mathbf{x}) &=& \sum b_{i_1,\dots,i_n} x_1^{i_1} \dots x_n^{i_n}.
\end{eqnarray*}
Then $I:=P_1/P_2$ is a rational integral of the ODE $\frac{d\mathbf{x}}{dt} = f(\mathbf{x})$ if
\begin{equation}
\dot{P_1}P_2 - P_1\dot{P_2} =0  \nonumber
\end{equation}
along solutions of the ODE. Here $\dot{}$ denotes $\frac{d}{dt}$.

\medskip \noindent For a polynomial ODE, the problem of finding $P_1$ and $P_2$, as posed, is bilinear in the parameters $a_{i_1,\dots,i_n}$ and $b_{i_1,\dots,i_n}$.

\subsection{Darboux polynomials (ODE case)}
\medskip \noindent Let $P(\mathbf{x})$ and $C(\mathbf{x})$ be polynomials.

\medskip \noindent Then $P(\mathbf{x})$ is called a Darboux polynomial of the polynomial ODE $\frac{d\mathbf{x}}{dt} = f(\mathbf{x})$, if
\begin{equation}
\dot{P}(\mathbf{x}) = C(\mathbf{x})P(\mathbf{x}).  \nonumber
\end{equation}
Here $C(\mathbf{x})$ is called the co-factor of $P$.

\medskip \noindent Note that $P(\mathbf{\mathbf{x}}(0))=0$ implies $P(\mathbf{x}(t))=0$ for all $t$. Hence the set $P(\mathbf{x})=0$ is an invariant set in phase space.

\medskip \noindent Consider two Darboux polynomials with the same co-factor $C$:
\begin{equation}
\begin{array}{c} 
\dot{P}_1 = CP_1 \\
\dot{P}_2 = CP_2 \end{array} 
\Rightarrow \frac{d}{dt} \left(\frac{P_1}{P_2}\right) = \frac{\dot{P}_1 P_2 - P_1 \dot{P}_2}{P_2^2} = \frac{CP_1P_2-P_1CP_2}{P_2^2} = 0,
\end{equation}
i.e. the ratio of two Darboux polynomials with the same cofactor is a rational integral. (The converse is also true).

\medskip \noindent However, finding $C$, $P_1$ and $P_2$ involves one bilinear problem, plus one linear problem. (Nevertheless, this approach can still be useful).

\medskip \noindent More generally,
\begin{equation}
\begin{array}{c} 
\dot{P}_1 = C_1P_1 \\
\dot{P}_2 = C_2P_2 \end{array} 
\Rightarrow \frac{d}{dt} (P_1P_2) = \dot{P}_1P_2 + P_1\dot{P}_2 = (C_1 + C_2)P_1P_2.
\end{equation}
A very nice introduction to Darboux polynomials for ODEs was given by Goriely \cite{G}. Note that Darboux polynomials were studied by Darboux, Poincar\'{e}, Painlev\'{e} and others \cite{G}, and are also known by several other names, including ``second integrals" and ``weak integrals".

\subsection{Discrete Darboux Polynomials (mapping case)}
Instead of polynomial ODEs $\frac{d\mathbf{x}}{dt} = f(\mathbf{x})$, we now consider rational maps $\mathbf{x}_{n+1} = \phi(\mathbf{x}_n)$ (cf \cite{FV,GM}).

\medskip \noindent Then we define $P(\mathbf{x})$ to be a Discrete Darboux Polynomial of the rational map $\mathbf{x}_{n+1} = \phi(\mathbf{x}_n)$ if
\begin{equation}
P(\mathbf{x}_{n+1}) = C(\mathbf{x}_n)P(\mathbf{x}_n),  \nonumber
\end{equation}
where the co-factor $C$ is now a rational function whose form will be presented in \S1.3.

\medskip \noindent We use the shorthand notation
\begin{equation}
P' = CP  \nonumber
\end{equation}
Note that, similarly to the continuous case, $P(\mathbf{x})=0$ is an invariant set in phase space.

\medskip \noindent Now consider again two Discrete Darboux Polynomials $P_1$ and $P_2$ with the same co-factor $C$:
\begin{equation}
\begin{array}{c} 
P_1' = CP_1 \\
P_2' = CP_2 \end{array} 
\Rightarrow \frac{P_1'}{P_2'} = \frac{P_1}{P_2},
  \nonumber
\end{equation}
i.e. the ratio of the two Discrete Darboux Polynomials with the same co-factors is again an integral (and the converse is also true).

\medskip \noindent More generally
\begin{equation}
\begin{array}{c} 
P_1' = C_1P_1 \\
P_2' = C_2P_2 \end{array}  
\Rightarrow (P_1 P_2)' = C_1C_2 (P_1P_2)
  \nonumber
\end{equation} 

\medskip \noindent How is all this going to help us find integrals of a given map?

\medskip \noindent  The answer comes in two parts:

\begin{enumerate}
\item In the discrete case we use a non-trivial ansatz for the co-factors $C(\mathbf{x})$. This ansatz works in all examples we have tried so far.
\item In the discrete case the co-factor of the product is the {\it product} of the co-factors. \newline In the continuous case the co-factor of the product is the {\it sum} of the co-factors.
\end{enumerate}

\medskip \noindent The latter point is crucial: It means that in the discrete case we can use the fact that the factorization of the co-factor $C$ is unique. By contrast, in the ODE case we have addition, where splitting into summands is not unique.
		
\subsection{Ansatz}

Ansatz: The co-factors we use are of the form
\begin{equation}
C(\mathbf{x}) = \frac{1}{D^l(\mathbf{x})} \prod_{i}^{}K_i^{a_i}(\mathbf{x}) \nonumber
\end{equation}
where $D(\mathbf{x})$ is the common denominator of the map, and the $K_i(\mathbf{x})$ are factors of the numerator of the Jacobian determinant  $J(\mathbf{x})$ of the map:
\begin{equation}
J(\mathbf{x}) = \frac{1}{D^m(\mathbf{x})} \prod_{i}^{}K_i^{b_i}(\mathbf{x}) \nonumber
\end{equation}
Comments:
\begin{enumerate}
\item There is a finite number of these co-factors up to a certain degree.
\item For each of this finite number of co-factors, we only need to solve a linear problem (up to a chosen degree).
\item If $C(\mathbf{x}) = J(\mathbf{x})$, the corresponding Darboux polynomials are (inverse) densities of preserved measures.
\end{enumerate}

\section{Determining preserved measures and first and second integrals of rational maps}
In this section we study the following two-dimensional ODE as an example:
\begin{eqnarray}\label{2DODE}
\frac{dx}{dt} &=& x(x + 6y - 3) \\
\frac{dy}{dt} &=& y(-3y - 2x + 3) \nonumber
\end{eqnarray} 
The Kahan-Hirota-Kimura (KHK) discretization of (\ref{2DODE}) reads (cf \cite{CMOQ,CMMOQ,HK,HP,KH,K,PPS})
\begin{eqnarray}\label{2DKHK}
x' &=& \frac{x(1+h(x+6y-3) + \frac{h^2}{4}(9-6x))}{D(x)} \\
y' &=& \frac{y(1+h(3-2x-3y) + \frac{9h^2}{4}(1-2y))}{D(x)} \nonumber
\end{eqnarray} 
where the common denominator $D(\mathbf{x})$ of the map is given by
\begin{equation}\label{2Ddenom}
D(\mathbf{x}) :=1 - \frac{h^2}{4}(9 - 12x - 36y + 4x^2 + 12xy + 36y^2)
\end{equation}
The Jacobian determinant $J(\mathbf{x})$ of the mapping (\ref{2DKHK}) is
\begin{equation}\label{2DJac}
J(\mathbf{x}) = \frac{K_1(\mathbf{x})K_2(\mathbf{x})K_3(\mathbf{x})}{D^3(\mathbf{x})}
\end{equation}
where
\begin{eqnarray}\label{2DKs}
K_1 &=& 1 + h(x-3y) - \frac{3}{4}h^2(3-2x-6y)  \nonumber \\
K_2 &=& 1 + h(x+6y-3) - \frac{3}{4}h^2(3-2x)   \\
K_3 &=& 1 + h(3-2x-3y) + \frac{9}{4}h^2(1-2y)  \nonumber
\end{eqnarray}
We have used cofactors $C_1=\frac{K_1}{D}$, $C_2=\frac{K_2}{D}$, $C_3=\frac{K_3}{D}$, $C_4=J$ to find the corresponding Discrete Darboux Polynomials for the map (\ref{2DKHK}):
\begin{align*}
p_{1,1} =& x+3y-3 \\
p_{2,1} =& x \\
p_{3,1} =& y  \\
p_{4,1} =& xy(x+3y-3) \\
p_{4,2} =& 1-\frac{h^2}{4}(9-12x-36y+4x^2+12xy+36y^2)
\end{align*}
Here $p_{i,j}$ denotes the $j^{th}$ Darboux polynomial corresponding to the cofactor $C_i$.

\medskip \noindent A phase plot for the map (\ref{2DKHK}), clearly exhibiting the linear Darboux polynomials $p_{1,1}$, $p_{2,1}$, and $p_{3,1}$, is given in Figure 1.

\begin{figure}[h!]
\begin{center}
\includegraphics[scale=.3]{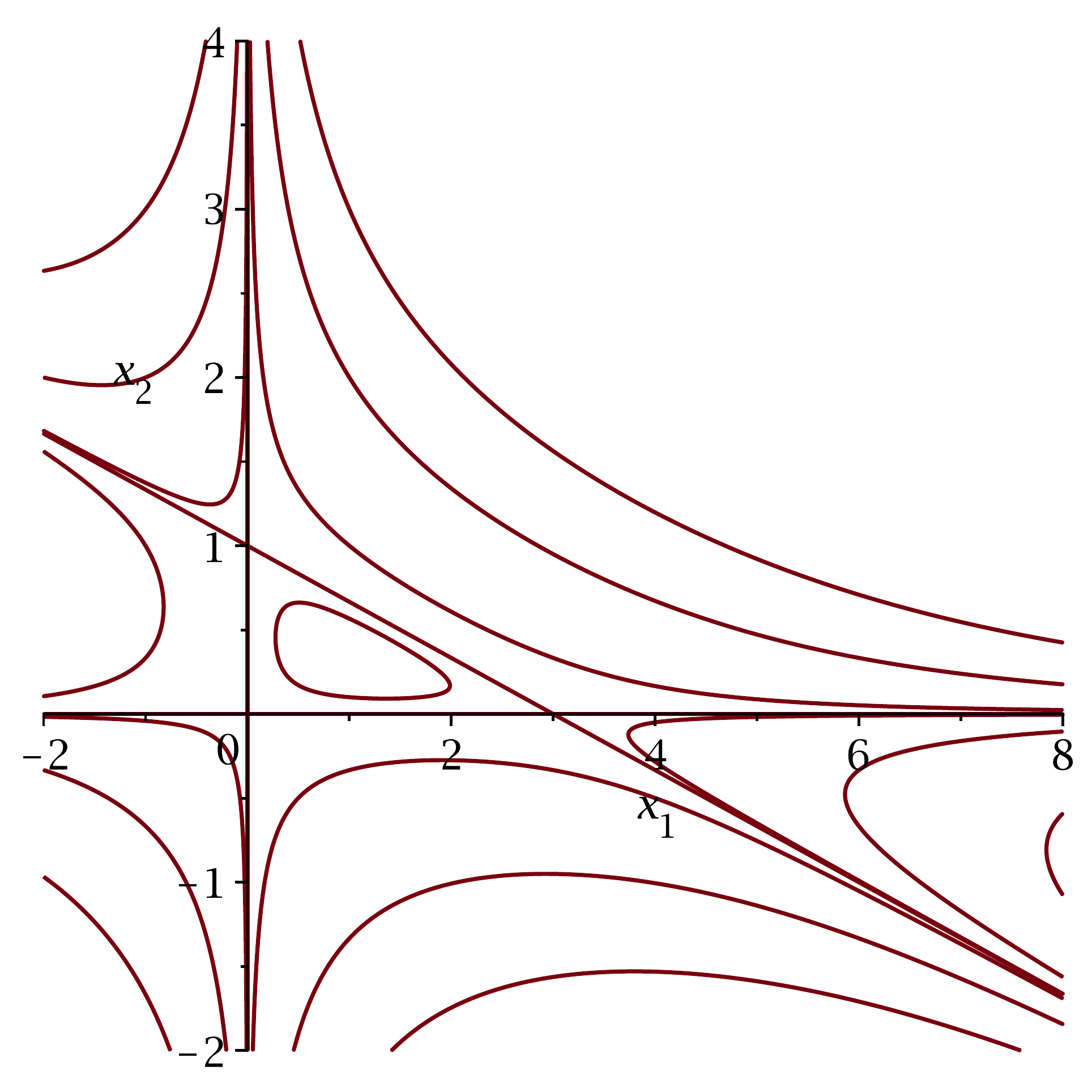}
\end{center}
\caption{Phase plot for map \eqref{2DKHK} and for $h=\frac{1}{17}$}
\end{figure}

\medskip \noindent It follows that the map (\ref{2DKHK}) preserves the integral
\begin{equation}\label{2Dint}
\tilde{I}(\mathbf{x}) = \frac{xy(x+3y-3)}{1-\frac{h^2}{4}(9-12x-36y+4x^2+12xy+36y^2)}
\end{equation}
and the measure
\begin{equation}\label{2Dmeas}
\frac{dxdy}{1-\frac{h^2}{4}(9-12x-36y+4x^2+12xy+36y^2)}
\end{equation}
Taking the continuum limit $h \rightarrow 0$, we obtain the cofactors $\tilde{C}_1 = x-3y$, $\tilde{C}_2 = x+6y-3$, $\tilde{C}_3 = 3-2x-3y$, $\tilde{C}_4 = 0$, and the corresponding Darboux polynomials
\begin{align*}
p_{1,1} =& x+3y-3 \\
p_{2,1} =& x  \\
p_{3,1} =& y  \\
p_{4,1} =& xy(x+3y-3) \\
p_{4,2} =& 1
\end{align*}

\noindent It follows that the ODE (\ref{2DODE}) preserves the integral
\begin{equation}\label{2DODEint}
I(\mathbf{x}) = xy(x+3y-3)
\end{equation}
and the measure
\begin{equation}\label{2DODEmeas}
dxdy.
\end{equation}
It thus turns out that our original ODE (\ref{2DODE}) is Hamiltonian, with $H(x)=xy(x+3y-3)$.

\medskip \noindent Interpreted conversely, one can say that the KHK discretization (\ref{2DKHK}) preserves the three affine Darboux polynomials of the ODE (\ref{2DODE}), as well as the modified integral (\ref{2Dint}) and the modified density (\ref{2Dmeas}). These results are no coincidences.

\medskip \noindent Indeed, the preservation of the three affine Darboux polynomials is the consequence of the following theorem (whose proof we will present elsewhere).
\begin{theorem}
The KHK discretization preserves all affine Darboux polynomials of a given quadratic ODE.
\end{theorem}

\medskip \noindent Theorem 1 is a very significant step towards the full resolution of the open problem posed in 2002 in \cite{MQ02}: `How does one preserve more than $n-1$ integrals and weak integrals (of an $n$-dimensional vector field)?'

\medskip \noindent The preservation of the modified integral and measure is an example of a general result in \cite{CMOQ} giving a modified integral for all systems with a cubic Hamiltonian in any dimension.

\section{Detecting preserved measures and first and second integrals of rational maps}
In this section we consider the following three-dimensional ODE as an example:
\begin{eqnarray}\label{3DODE}
\frac{dx}{dt} &=& x(y-\mu z)  \nonumber \\
\frac{dy}{dt} &=& y(\lambda z - x)   \\
\frac{dz}{dt} &=& z(\mu x - \lambda y)  \nonumber,
\end{eqnarray}
where $\lambda$ and $\mu$ are arbitrary parameters.

\medskip \noindent Applying the Kahan-Hirota-Kimura discretization to (\ref{3DODE}), we obtain
\begin{eqnarray}\label{3DKHK}
\frac{x'-x}{h} &=& \frac{x'(y-\mu z) + x(y'-\mu z')}{2}  \nonumber \\
\frac{y'-y}{h} &=& \frac{y'(\lambda z - x) + y(\lambda z' - x')}{2}   \\
\frac{z'-z}{h} &=& \frac{z'(\mu x - \lambda y) + z(\mu x' - \lambda y')}{2}  \nonumber. 
\end{eqnarray}
Solving equation (\ref{3DKHK}) for $x'$, $y'$,  and $z'$ we obtain the (rational) Kahan map discretizing (\ref{3DODE}). Using the Jacobian determinant $J(\mathbf{x})$ of the Kahan map as cofactor, our algorithm finds that for all $(\mu,\lambda)$, the map preserves the measure $\frac{dx dy dz}{xyz}$ and the first integral $x+y+z$.

\medskip \noindent Moreover, the algorithm also detects the following special values of the parameters $(\mu,\lambda)$ where the map preserves an additional integral, and outputs the formula for the integral (cf. Table 1).

\begin{table}[h!]
  \begin{center}
    \caption{Integrable parameter values  and corresponding functionally independent additional first integrals detected by our algorithm.}
    \label{tab:table1}
    \begin{tabular}{c|l}
       \hline
      $(\mu,\lambda)$ & additional first integral  \\
      \hline
      $(-1,0)$ & $y/z$ \\
      $(1,0)$ & $yz/(1-\frac{h^2}{4}x^2)$ \\
      $(0,1)$ & $xz/(1-\frac{h^2}{4}y^2)$ \\
      $(0,-1)$ & $z/x$ \\ 
      $(1,1)$ & $xyz/(1-\frac{h^2}{4}(x^2 + y^2 + z^2 -2xy - 2xz -2yz))$ \\ 
      $(1,-1)$ & $x/yz$ \\ 
      $(-1,-1)$ & $z/xy$ \\ 
      $(-1,1)$ & $y/xz$ \\ 
      \hline
    \end{tabular}
  \end{center}
\end{table}

\section{Concluding remarks}
In this Letter we have presented a method for detecting and determining first and second integrals of rational maps. There are in the literature several other methods for {\it determining} first and second integrals of discrete systems, cf. \cite{FV,GM,PNC,TKQ} and references therein. There are also in the literature several other methods for {\it detecting} first and second integrals of discrete systems, cf. \cite{AABM,HK,RV} and references therein.

\medskip \noindent However, to our knowledge none of the above combine all the following properties of the method  presented in this Letter:
\begin{enumerate}
\item It is algorithmic, and requires no other input than the rational map in question. At heart the algorithm is linear and, to some extent apart from birationality, requires no knowledge about the map (such as symplecticity, measure preservation, time-reversal symmetry, integrability, Lax pairs, etc) on the part of the user.
\item Up to a certain prescribed degree, it determines and outputs all:
  \begin{enumerate}
  \item rational first integrals
  \item polynomial second integrals
  \item preserved measures of the form $P(x) dx$ or $\frac{dx}{P(x)}$, where $P$ is a polynomial.
  \end{enumerate}
\item It can detect special parameter values where additional preserved first and/or second integrals and/or measures exist, and output those integrals and measures.
\item It works for both integrable and non-integrable cases.
\item It allows one to take the continuum limit, if appropriate.
\end{enumerate}


\section*{Acknowledgements}
This work was supported by the Australian Research Council, by the Research Council of Norway, and by the European Union's Horizon 2020 research and innovation program under the Marie Sk\l{}odowska-Curie grant agreement No. 691070. GRWQ is grateful to K.Maruno for his hospitality at Waseda University, and to G.Gubbiotti for useful comments and correspondence.


\end{document}